\documentclass[10pt]{article}

\usepackage[ansinew]{inputenc}
\usepackage[T1]{fontenc}
\usepackage{latexsym}
\usepackage{graphicx}
\usepackage{color}
\usepackage{natbib}
\usepackage{SIunits}

\newcommand{\up}[1]{\raisebox{1ex}{\tiny #1}}

\newcommand{\fsiz}[0]{0.6\textwidth}

\newcommand{\mr}[0]{\hline{}}
\newcommand{\br}{\hline{}}
\newcommand{\0}{ }

\newcommand{\uwlf}[1]{\ensuremath{\mbox{ W(m K)\up{-1}}}}
\newcommand{\utd}[1]{\ensuremath{\mbox{ m\up{2} s\up{-1}}}}
\newcommand{\uhf}[1]{\ensuremath{\mbox{ W m\up{-2}}}}

\begin{document}

\bibliographystyle{agsm}

\title{Uncertainties and
  shortcomings of ground surface temperature histories derived from
  inversion of temperature logs}
\author{A Hartmann and V Rath\\
  Applied Geophysics, RWTH Aachen University }
\date{September 29, 2005}

\maketitle{}
\begin{abstract}
  Analysing borehole temperature data in terms of ground surface
  history can add useful information to reconstructions of past
  climates. Therefore, a rigorous assessment of uncertainties and
  error sources is a necessary prerequisite for the meaningful
  interpretation of such ground surface temperature histories. This
  study analyses the most prominent sources of uncertainty.  The
  diffusive nature of the process makes the inversion relatively
  robust against incomplete knowledge of the thermal diffusivity.
  Similarly the influence of heat production is small. It turns out
  that for investigations of the last 1000 to 100\,000 years the
  maximum depth of the temperature log is crucial. More than 3000~m
  are required for an optimal inversion.  Reconstructions of the last
  one or two millennia require only modestly deep logs (>300~m) but
  suffer severely from noisy data.
\end{abstract}


\emph{submitted to Journal of Geophysics and Engineering}

\section{Introduction}
\label{sec:intro}

Borehole temperature data record past changes in ground surface
temperature (GST) as the transient temperature signal diffuses into
the subsurface. These temperature data have been used to reconstruct
time series of the ground surface temperature in numerous studies.
The value of such reconstructions as a proxy for the paleoclimate,
however, is still subject of research and discussion
\citep[e.g.][]{Mann2003a}. It has been shown
\citep[e.g.][]{Gonzalez-Rouco2003,Signorelli2004} that surface air
temperature (SAT) and soil temperature correlate well at long
timescales.  Reconstructions have been carried out on various
timescales from a few hundred to 100\,000 years by several authors.
While the shorter type serves, combined with proxy climate data, as a
means to investigate natural and anthropogenic climate change, the
longer reconstructions study temperatures of the last ice-age and its
transition to our current climate.

The problem is usually simplified to a purely conductive,
one-dimensional one. If the model fails to capture reality, any
inverse method will produce spurious results. Previous studies have
already analyzed various effects that can possibly influence the
transient signal. For instance, the error introduced by the common
assumption of one-dimensionality has been treated by
\citet{Shen1995} and \citet{Kohl1999}. Advection of heat by
groundwater flow can modify the signal considerably
\citep{Clauser1997,Kohl1998,Taniguchi1999a,Reiter2005}.  In the
following analysis we will show that even with all simplifying
assumptions fulfilled, it is possible to produce artefacts in the
inverted time series. Although we use a particular algorithm for the
study, the results will more or less apply to other programs as
well. We will start with a short review of the theory of the
algorithm in the next section.

\subsection{Theory}
\label{sec:Theory}

The algorithms commonly used for GST history inversion assume a
one-dimensional, purely conductive model.  The physical properties of
the subsurface are known, the medium can be layered
\citep{Shen1991,Shen1992} or is assumed to be homogeneous
\citep{Mareschal1992,Beltrami1995b}. Both versions can be used on
single boreholes or on ensembles of temperature logs in order to
improve the signal to noise ratio.  In a comparison of different codes
\citep{Beck1992b}, it was found that all of them perform equally well
under similar circumstances.

In the model considered here \citep{Mareschal1992,Beltrami1995b},
the subsurface is considered homogeneous. The steady state
temperature profile is defined by the heat flux at the surface
$q_0$, the pre-observational mean ground surface temperature $T_0$,
thermal conductivity $\lambda$, and heat production rate $A$ of the
subsurface. Together with a additional transient term $T_t(z,t)$ the
temperature at time $t$ and depths $z$ is given by
\citep{Carslaw1959,Beltrami1995b}:
\begin{equation}
  \label{eq:homogenousforwardmodel}
  T(z,t) = T_0  + \frac{q_0z}{\lambda} - \frac{Az^2}{\lambda} + T_t (z,t).
\end{equation}
In the particular case that the GST history can be parameterized by
stepwise temperature changes $T^G_j$ at times $t_j$ before present
($t=0$) the transient term becomes
\begin{equation}
  \label{eq:forwardmodel2}
  T_t(z)  = \sum\limits_{j = 1}^N
  {T_n^G \left( {\mbox{erfc} \left( {\frac{z} {{2\sqrt
                {\kappa t_j } }}} \right) - \mbox{erfc} \left( {\frac{z}
            {{2\sqrt {\kappa t_{j - 1} } }}} \right)} \right)}
\end{equation}
If thermal conductivity, diffusivity and heat production are known,
the only remaining unknowns are the steady-state GST, the surface heat
flow and the values of the the GST history. If temperatures are
recorded at depths $z_i \quad i=1,...,M$, a set of M linear equations
can be formed from equation~\ref{eq:homogenousforwardmodel}, that can
be solved for the unknown variables. The solution needs to be
regularized because the problem is ill-posed. The singular value
decomposition \citep{Lanczos1961} is used for this purpose by solving
the linear problem and at the same time seeking a solution that
minimizes the norm of the model. Implications of this procedure will
be discussed in more detail in section~\ref{sec:l-curve}.

The original code\footnote[1]{A MATLAB version of the program
  including a GUI and the optimal choice of the regularization
  parameter by
  means of the L-curve method can be  downloaded from:
  \texttt{http://www.geophysik.rwth-aachen.de/Downloads/software.html}}
was employed in various case studies for the reconstruction of past
climates
\citep[e.g.][]{Beltrami1997,Beltrami2004,Clauser1995a,Clauser1997,Jones1999}.
It was also used in conjunction with other proxy data to improve the
temporal resolution of the GST history as well as the coupling between
GST and SAT
\citep{Beltrami1995a,Beltrami1995c,Signorelli2004,Nitoiu2005}.  Before
analyzing the sensitivity of this algorithm it is worthwhile to
investigate the effect of changing GST on the temperature measured in
boreholes.

\subsection{Paleoclimate influence on subsurface temperatures}
\label{sec:PaleoInfluence}

\begin{figure}
  \begin{center}
    \includegraphics[width=\fsiz,clip,trim=0 0 0 0]{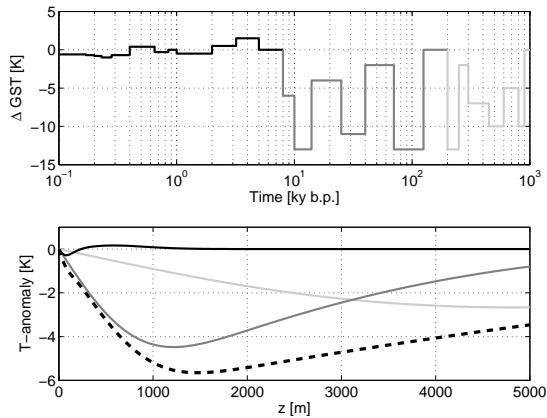}
    \caption[Time and depth scales of  transients in borehole
    temperatures due to paleoclimate ]
    {Time and depth scales of GST changes and respective changes in
      today's subsurface temperature. Top: Idealized GST history
      \citep[table 10.9]{Haenel1988}. The different line styles mark
      characteristic periods, respectively. Bottom: Temperature anomalies in
      the subsurface computed from the time series. Line colours of
      the anomalies correspond to the periods of the upper panel.
      The dashed line is equivalent to the sum of the anomalies.}
    \label{fig:timescales}
  \end{center}
\end{figure}
The magnitude and time/depth scale of GST changes can be seen in
figure~\ref{fig:timescales}. We used a temperature history of the last
1 million years based proxy data \citep[table 10.9]{Haenel1988} to
demonstrate the effect of this transient signal on the subsurface
temperatures. The curve is valid only for Switzerland but it displays
the main features of the climate of the last 1 million years.  The
proxy curve is subdivided into 3 parts and for each a perturbation
with respect to the mean is calculated. The transient signals for each
of the periods are computed separately (figure~\ref{fig:timescales},
lower panel, solid lines). The mean GST ($T_0$ in
equation~\ref{eq:homogenousforwardmodel}) for each of the computations
is assumed to be 0°C.

The total transient signal (dashed line) is then the sum of the three
partial signals.  It is apparent that changes of the order of
$10^6$~years for this particular GST history will appear in a
temperature log as a virtually straight line with a slope of about
-1~K km\up{-1} unless the temperature log runs down to a depth of at
least 4000 m. The slope will change if a different GST history is used
but it is clear that this part of the transient signal cannot be
resolved by an inversion, especially not in the presence of noise.
Thus, fluctuations of the period 100\,000 to 1000\,000 years will
usually appear only in the pre-observational mean (POM). The glacial
temperature minimum and the postglacial temperature increase (100\,000
to 10\,000 years) cause the most prominent disturbance of the
temperature profile. The temperature deviation reaches its maximum
amplitude at about 1500~m depth, but still has a value of 1.5~K at
4000~m depth. Because most of the available undisturbed temperature
data are from smaller depths, a large portion of this signal is
frequently not recorded.  The signal due to temperature variations
during the Holocene can be traced down to a maximum depth of 1000~m.
Although temperature data in this depth range are more numerous,
reconstructions have been attempted using substantially shallower
data.

From the discussion above two questions arise.
Section~\ref{sec:l-curve} focuses on the question what the value of
the regularizing parameter $\epsilon$ for any given problem should be
and whether some objective criterion can be used to define it.  In
section~\ref{sec:sensitivity_analysis} the question is addressed if a
reliable reconstruction of past temperatures is possible when only
part of the transient signal is contained in the data.

\section{Optimum choice of regularization parameter}
\label{sec:l-curve}

It was mentioned in the introduction that a regularization is needed
to solve the inverse problem in a stable manner. For the singular
value decomposition, two methods are commonly used \citep{Menke1989}.
Either only the $p$ largest singular values are used in the inversion
or a damping parameter $\epsilon$ is added to the singular values such
that
\begin{equation}
  \label{eq:sv_plus_eps}
  \frac{1}{\lambda_i} = \frac{\lambda_i}{\epsilon^2 + \lambda_i^2}.
\end{equation}
The stabilization of the solution is obtained at the cost of loss of
model resolution. Both methods require a choice of a regularization
parameter $\beta$, i.e. the number of singular values discarded, or
the magnitude of the damping parameter.  A regularization of this
type tends to minimize the norm of the model. This is equivalent to
minimizing the objective function $\phi(m)$ given by
\citep{Aster2004,Farquharson2004}:
\begin{equation}
  \label{eq:general_norm}
  \phi(\mathbf{m}) = ||\mathbf{G}(\mathbf{m}) - \mathbf{d} ||
  + \epsilon^2 ||\mathbf{m}||.
\end{equation}
Here the first term represents the $L_2$-norm of the data misfit, with
$\mathbf{G(m)}$ being the forward model and $\mathbf{d}$ being the
data.  The second term represents the $L_2$-norm of the model
$\mathbf{m}$. In terms of paleoclimate inversion, minimizing the
model norm can possibly change the amplitude of paleoclimatic
disturbances. Too small an $\epsilon$, on the other hand, will lead
to spurious oscillations in the solution of the GST.  Thus it is
important to choose a proper value for the damping parameter,
preferably based on some objective criterion.

\begin{figure}
  \centering
  \includegraphics[width=.6\columnwidth,trim=0 20 0 0,clip]
  {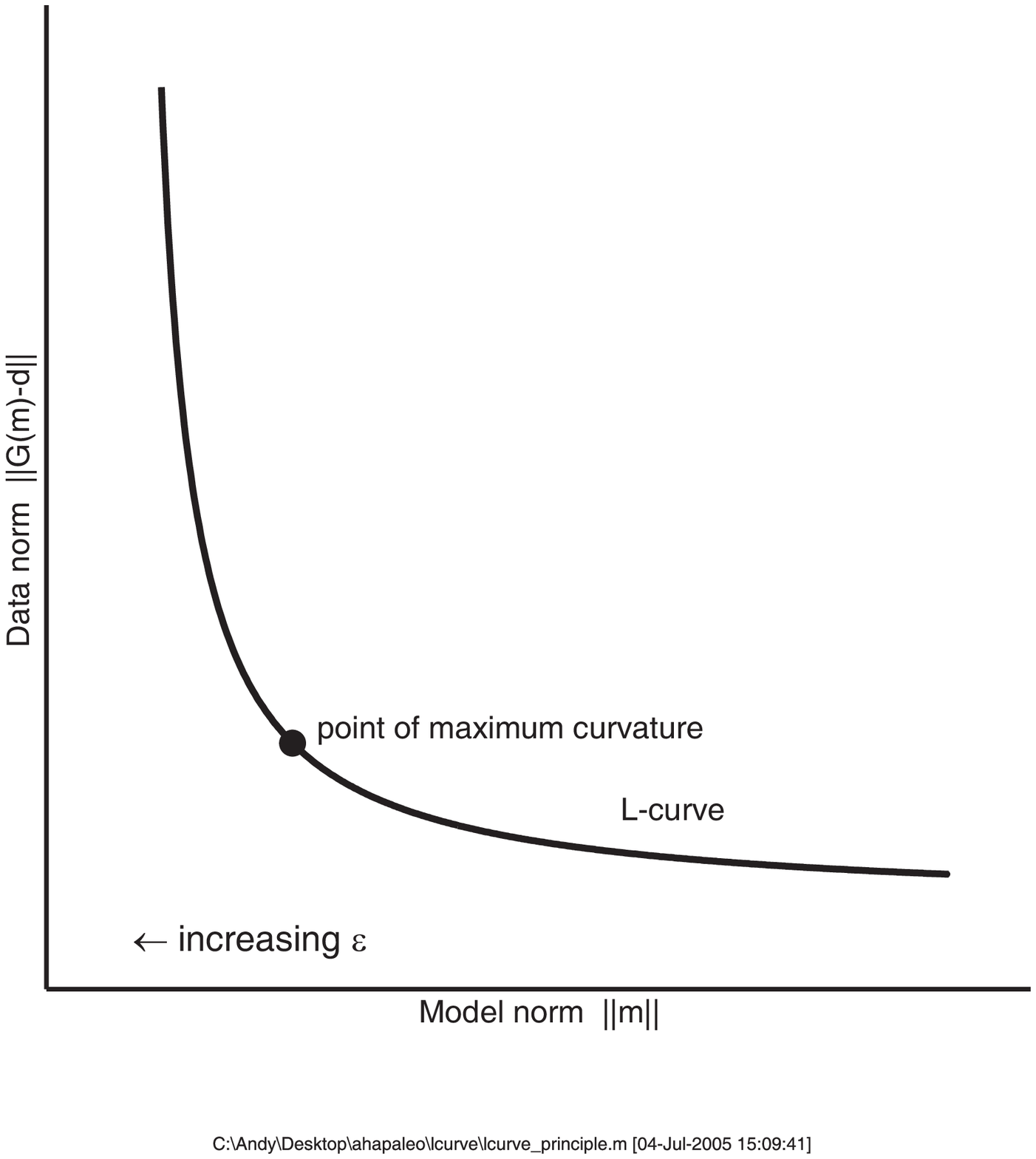}
  \caption[L-curve principle]{Principle of the L-curve. With
    increasing $\epsilon$ the model norm decreases and the data norm
    increases. The optimum value for $\epsilon$ can be found at the
    maximum curvature point of the curve.}
  \label{fig:lcurve_principle}
\end{figure}
One way to gain insight into the problem is the L-curve
\citep{Hansen1998,Farquharson2004}, so called for its shape. To create
this curve, the inversion is repeated for a range of regularization
parameters and the data norm is plotted versus the model norm
(figure~\ref{fig:lcurve_principle}) on a logarithmic scale. For large
values of $\epsilon$ the model norm is small but the data norm is
large. A small $\epsilon$ leads to a large model norm and to a data
misfit norm that is close to the variance of the data. These two cases
define the asymptotic behaviour of the L-curve. At the maximum
curvature point of the curve, a compromise is found between the
magnitude of those two norms \citep{Hansen1998}. This is the optimal
value for the regularization parameter.

\begin{table}[htbp]
  \caption[Parameters of the L-curve model]{Parameters of the synthetic
    model used for the L-curve studies.}
  \label{tab:lcurve_synth_parm}
  \begin{center}
\begin{tabular}{@{}lll}
\br
 Parameter &      Value &  Unit           \\
\mr
Sampling rate &     \0\020 &          m \\

Total depth of log &       5000 &          m \\

Number of time steps for inversion &     \0\020 &            \\

Start time of inversion &       1000 &    ky b.p. \\

Stop time of inversion &  \0\0\00.1 &    ky b.p. \\
\br
\end{tabular}

  \end{center}
\end{table}
We studied our inversion problem with the help of the L-curve
criterion. A synthetic temperature log was created using the parameter
set given in table~\ref{tab:lcurve_synth_parm}. The transient
temperature anomaly given in figure~\ref{fig:timescales} was added to
the steady state solution and random, normally distributed noise was
added. The noise amplitudes for the test data sets were 0.01, 0.2,
0.4, and 0.6 K, respectively.

\begin{figure}
  \begin{center}
    \includegraphics[width=\fsiz{},trim=0 10 0 0,clip]
    {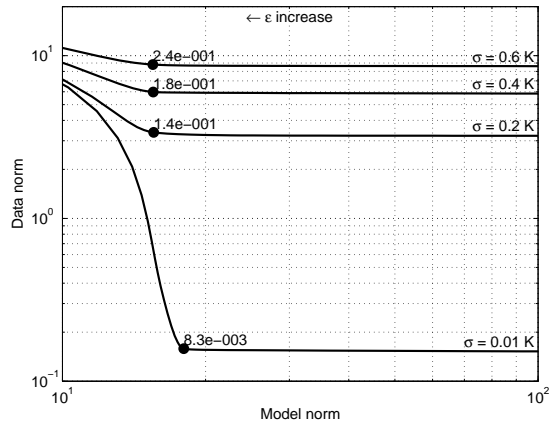}
  \end{center}
  \caption[L-Curve for different noise levels]{L-Curve for different
    noise levels.}
  \label{fig:lcurve_noise_level_opt_eps}
\end{figure}
\begin{figure}
  \begin{center}
    \includegraphics[width=\fsiz{},trim=0 20 0 0,clip]
    {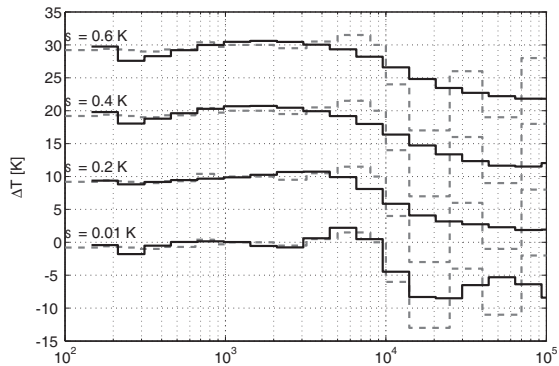}
  \end{center}
  \caption[Inversion results for different noise levels and optimal
  $\epsilon$]{Inverted GST-history for different noise levels and
    optimal $\epsilon$ from
    figure~\ref{fig:lcurve_noise_level_opt_eps}. The curves are each
    shifted by 5°C for better visibility.}
  \label{fig:lcurve_noise_level_gst}
\end{figure}

Figure~\ref{fig:lcurve_noise_level_opt_eps} shows the variation of the
L-curve and the optimum $\epsilon$ as a function of the noise level.
It is apparent that only for the smallest noise level the L-curve
takes on a well shaped form with an maximum curvature point easy to
determine. For all noise levels the L-curve displays a straight line
region where any change in $\epsilon$ leads to large changes in the
model norm but does not change the data fit at all. This means that a
value of $\epsilon$ less than the value found in the maximum curvature
point is not justified by the data. It is noteworthy that the model
norm of the optimal model does not vary much with the increasing noise
level.  This indicates that the inverted model is robust against noise
for this configuration. This point is illustrated by comparing the
inverted GST histories for optimal values of $\epsilon$
(figure~\ref{fig:lcurve_noise_level_gst}).
The major feature is the temperature increase at the end of the last
ice age. It is preserved in all of the results in comparable
magnitude and robust against noise. Smaller temperature variations
during the Holocene period, on the other hand, are only resolved at
the smallest noise level. The L-curve provides a tool to answer the
important question if a sought for feature in the paleoclimate
record can be resolved by the data.

\begin{figure}
  \centering
  \includegraphics[width=\fsiz,clip,trim=0 20 0 0]{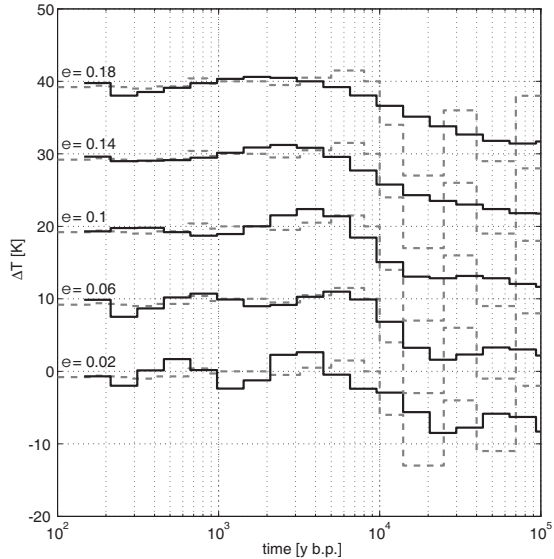}
  \caption[GST-histories for a fixed noise level and varying values of
  the damping parameter]{GST-histories for a fixed noise level of 0.2
    K and varying values of the damping parameter $\epsilon$. The
    optimal choice of $\epsilon$ according to the L-curve criterion is
    0.14.}
  \label{fig:lcurve_gst_many_eps}
\end{figure}
Finally, figure~\ref{fig:lcurve_gst_many_eps} shows a comparison of
inversion results for a given noise level and a range of
$\epsilon$-values. The optimal value would be about 0.14, yielding a
rather smooth model. The interpreter, possibly aware of a
paleoclimatic proxy curve like the ones shown in the figure, might be
tempted to choose a smaller value that seems to better fit the
temperature variations in the model during the Holocene. However,
figure~\ref{fig:lcurve_noise_level_opt_eps} indicates that the data
fit does not improve at all by choosing an $\epsilon$ smaller than
0.14. The small amplitude variations on top of the temperature
increase at the end of the ice age appear to be only numerical
artifacts.

\section{Sensitivity analysis}
\label{sec:sensitivity_analysis}

\subsection{Length of the temperature log}
\label{sec:SensLogLength}

\begin{figure}
  \begin{center}
    \includegraphics[width=\fsiz{},trim=0 10 0 0,clip]
    {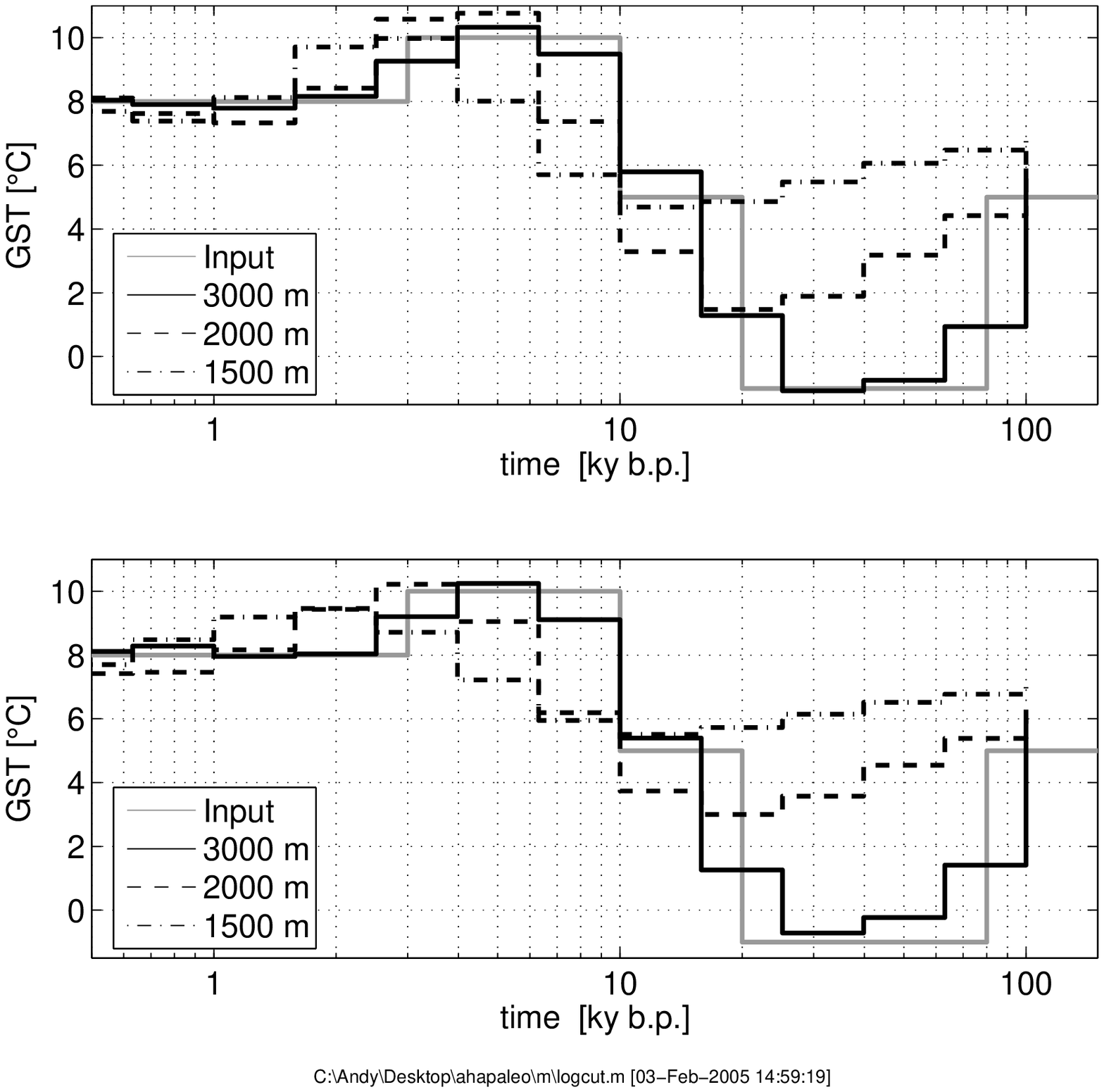}
  \end{center}
  \caption[Inversion results for shortened logs]{A shortening of logs
    results in different GST reconstructions. Solid line represents
    the GSTH that was used to drive the forward model. The dashed
    lines show reconstructions for different depths of the logs. The
    temperature of the last glacial minimum is only resolved for the
    3000~m log. Not even the timing of the climatic optimum in the
    Holocene is correctly resolved for the 1500~m log.}
  \label{fig:LogCut}
\end{figure}

\begin{table}[htbp]
  \caption{Values for the thermal properties that are  used throughout
    the text to compute  synthetic temperature-depth profiles.}
  \label{tab:thermal_params}
   \begin{center}
\begin{tabular}{lll}
\br
 Parameter &      Value &       Unit \\
\hline
Thermal conductivity &    $\02.5$ & W (m K)\up{-1} \\

Volumetric heat capacity & $\02.30\cdot10^6$ & J m\up{-3} K\up{-1} \\

Thermal diffusivity & $\01.09\cdot10^{-6}$ & m\up{2} s\up{-1} \\

Heat production rate & $\01.0\cdot10^{-6}$ & W m\up{-3} \\

Surface temperature &     $10.0$ &         °C \\

 Heat flow &     \00.06 & W m\up{-2} \\

Sampling rate &      \01.0 &          m \\
\br
\end{tabular}

  \end{center}
\end{table}

In the introduction the question was posed if the full depth range of
the transient signal is needed for a valid reconstruction of the
GST history. To analyze this problem, a
synthetic temperature log with a simplified GST
history of the last 1 million years is computed.  The parameters for
the temperature log are taken from table~\ref{tab:thermal_params}. The
GST history that produces the transient temperature signal (grey line,
figure~\ref{fig:LogCut}) does not represent an actual paleoclimatic
curve but rather represents an idealized history with correct order of
magnitude for amplitude and timing of the events.

For the inversion procedure, the originally 3000~m deep log was cut off
at depths of  2000~m and 1500~m. All three logs were inverted with and
without synthetic noise applied. We chose  normally distributed noise
with a standard deviation of 0.3~K. For the inversion of the noise
free and noisy logs, a constant damping parameter $\epsilon$ of 0.1
and 0.2, respectively, was used.

The resulting GST histories for the different inversion runs are
displayed in figure~\ref{fig:LogCut}. For both the noisy and the noise
free data a log depth of 3000~m seems to be sufficient to fully
recover the amplitude of the last glacial stage. A decrease of the
maximum log depth also reduces the amplitude of the glacial cooling.
Further the phase of the events is shifted to later times. The effect
is more pronounced for the noisy data, but even for noise free data
any reconstruction of the last ice age paleoclimate based on logs
shallower than 2000~m seems to be problematic.

\begin{figure}[tbp]
  \centering
  \includegraphics[clip,trim=0 10 0 0,width=.8\columnwidth]
  {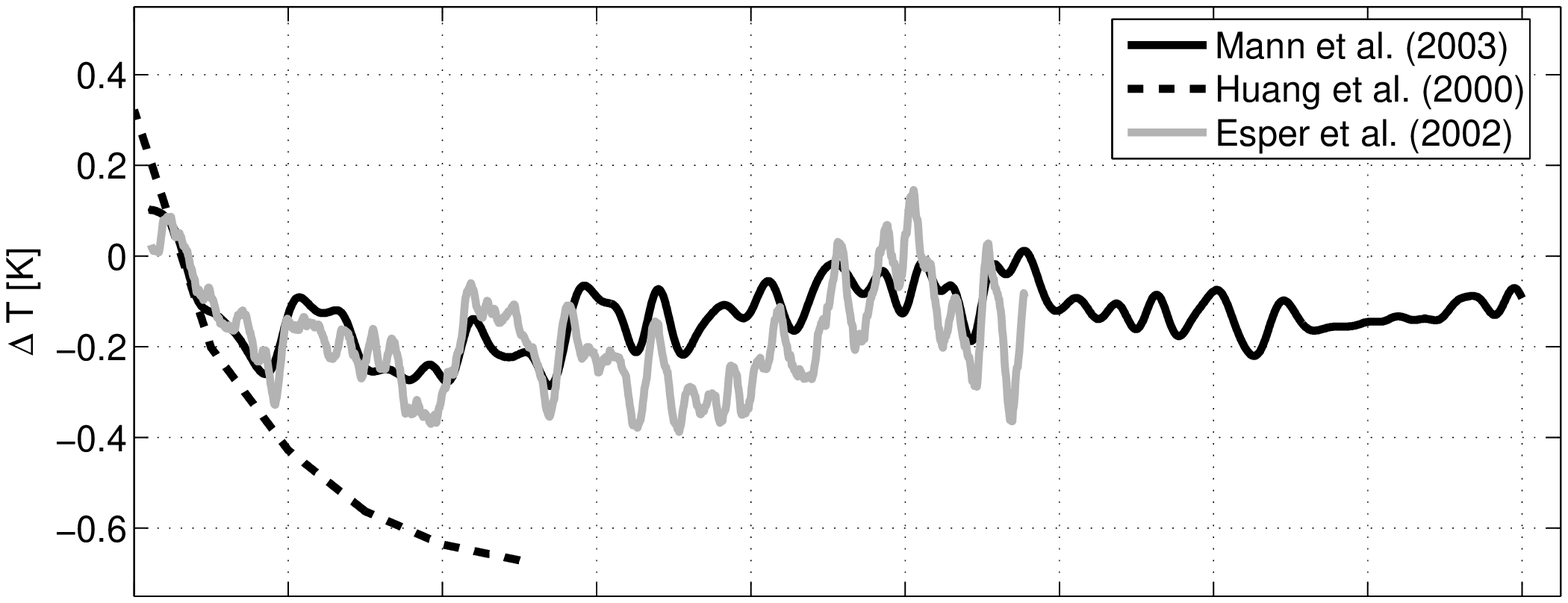}
  \includegraphics[clip,trim=0 0 0 5,width=.8\columnwidth]
  {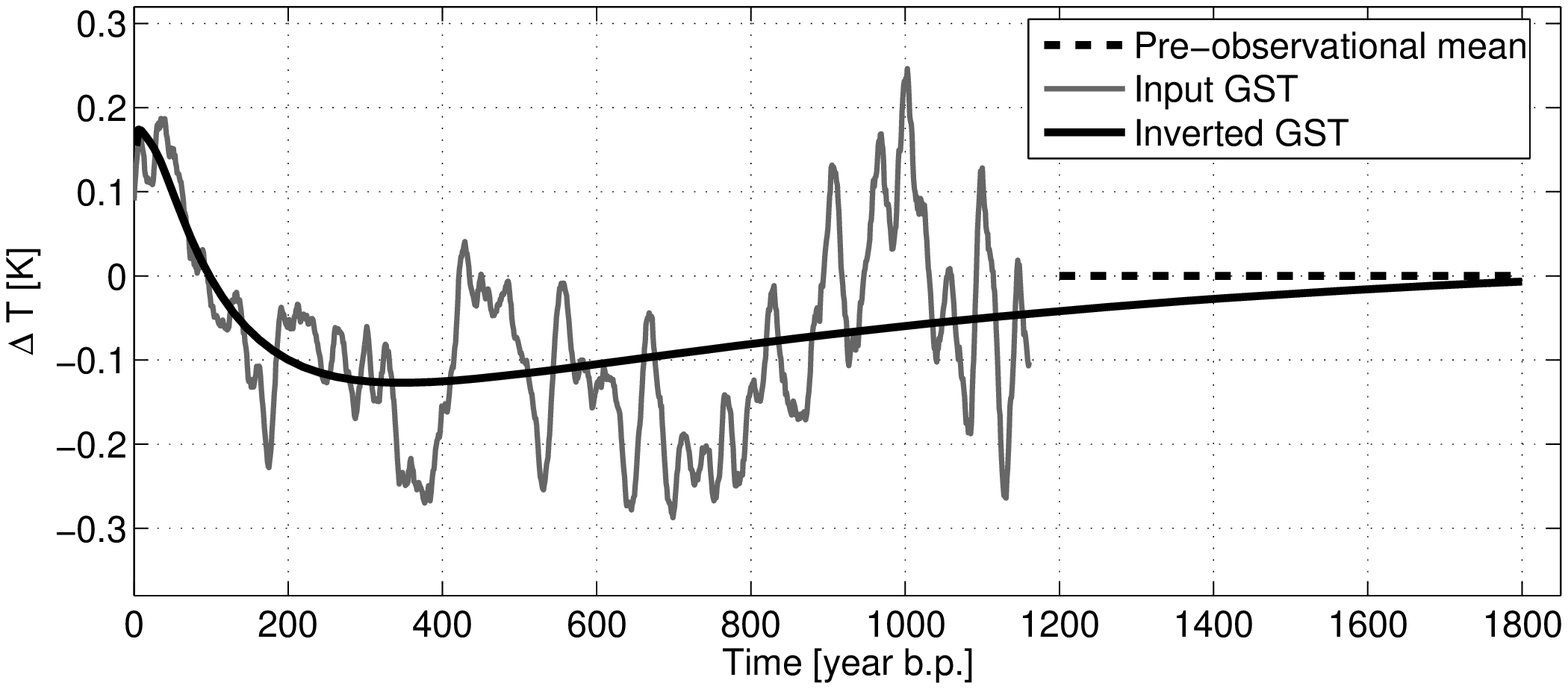}
  \caption{Upper panel: Various reconstructions of paleo temperatures
    of the last two millennia. Lower panel: Inversion results for a
    2000 m deep noise free temperature log using the proxy-curve of
    \citet{Esper2002} as input data.}
  \label{fig:mill_inv_long_log}
  \label{fig:proxy_comparison}
\end{figure}
A similar procedure of shortening logs can be used to estimate the
necessary depth of temperature logs for millennial scale climate
reconstructions. Currently plenty of proxy reconstructions for the
last one or two millennia exist and their validity is debated. The
geothermal method can provide valuable insight into this problem as it
is the only direct measurement of paleo-temperatures. The more it is
important to analyze the information contained in the temperature
data.  Figure~\ref{fig:proxy_comparison}, upper panel, shows a few
reconstructions of the Northern hemisphere annual mean SAT together
with a reconstruction based on borehole temperature data
\citep{Mann2003b,Esper2002,Huang2000}.

In our analysis we use the reconstruction of \cite{Esper2002} as the
forcing to compute the transient temperature perturbation. The other
parameters of the forward model are taken again from
table~\ref{tab:thermal_params}.  Additionally, a pre-observational
mean needs to be specified. As the proxy curve contains no information
about this parameter, it was arbitrarily chosen to be the mean
temperature in the interval 800--1200 years before present.
Figure~\ref{fig:mill_inv_long_log}, lower panel, displays an inversion
result for this synthetic temperature log. The temperature data are
noise free and the log runs 2000 m deep.  Because of the favourable
conditions for this particular inversion run the information contained
in this reconstruction can be considered the maximum knowledge that
can be extracted by inverting temperature-depth data. The warming
trend of the last 200 years is captured very well. No insight into
climate variations of the preceding centuries can be gained, but the
pre-observational mean is preserved in the reconstruction. Thus the
borehole method effectively provides two unknown pieces of
information: The temperature change from the average temperature of
the last millennium and the mean temperature before that period.

\begin{figure}[htbp]
  \centering
  \includegraphics[width=.6\columnwidth,clip,trim=0 32 0 0]
  {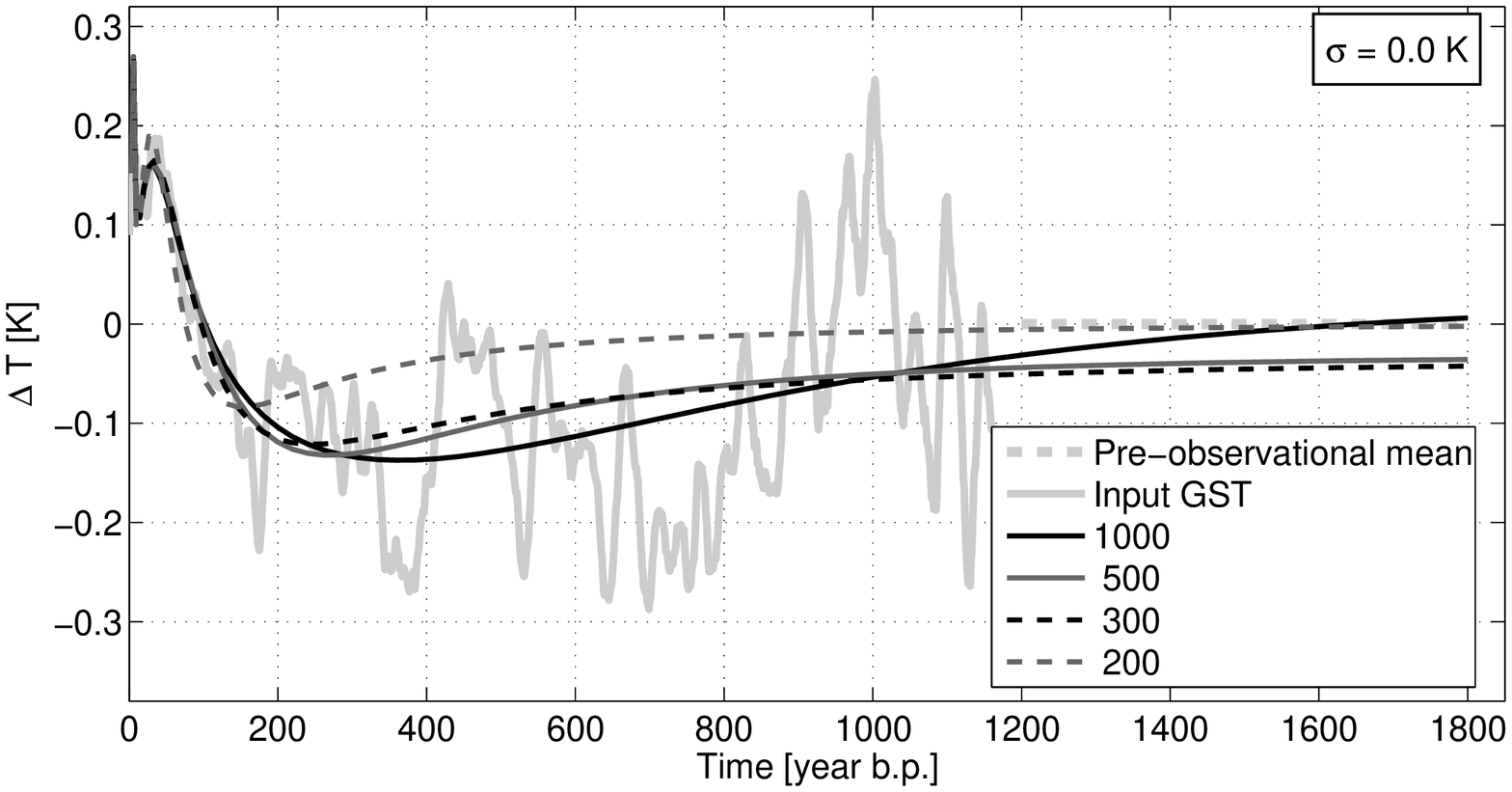}
  \includegraphics[width=.6\columnwidth]
  {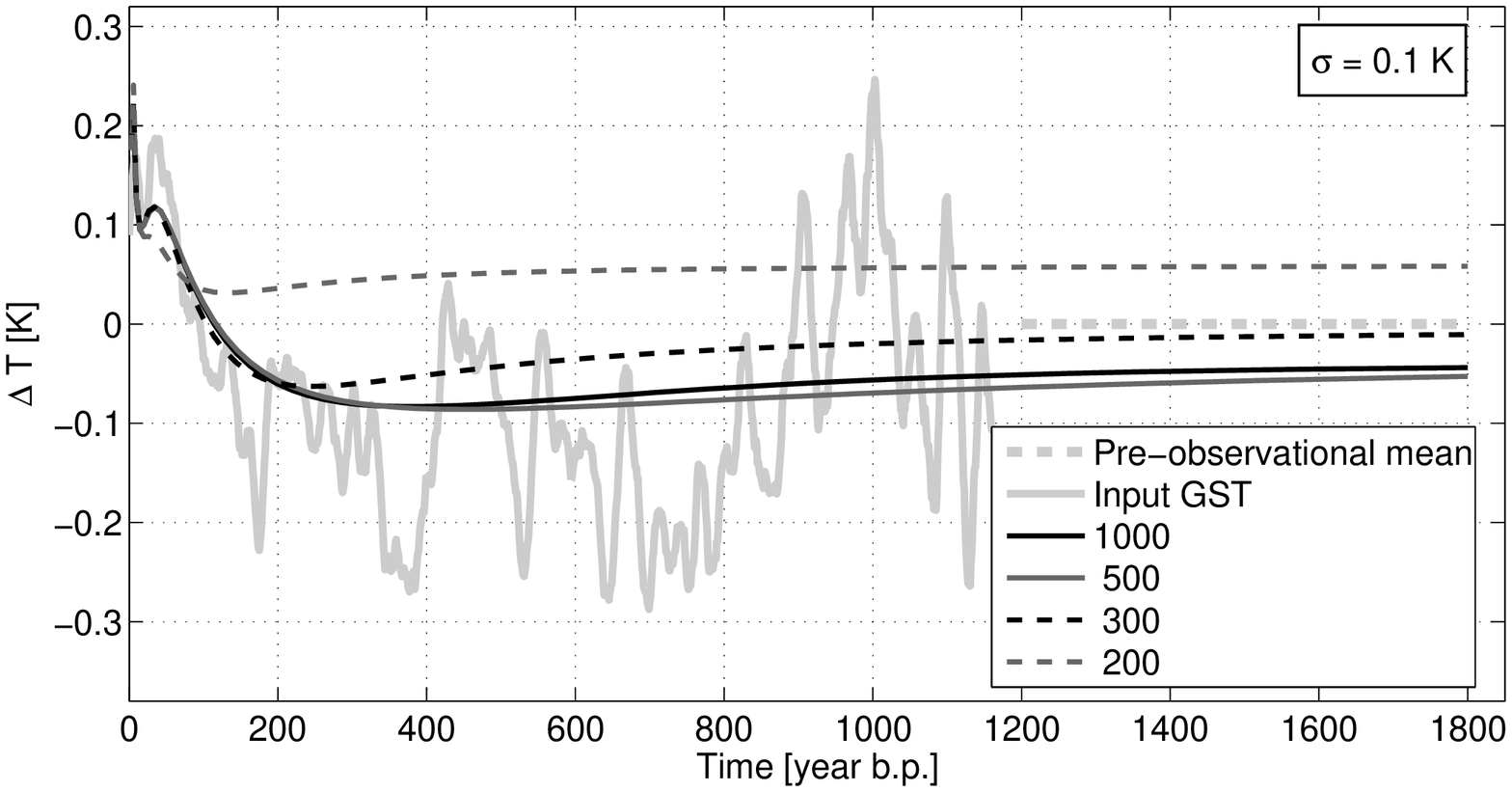}
  \caption{Inversion results for decreasing maximum depth of the
    temperature log. Temperature logs were cut off at various depths.
    Thermal parameters are the same as in
    table~\ref{tab:thermal_params}. Upper panel: Input temperature
    data for the inversion was noise-free. The damping parameter
    $\epsilon$ equals 0.1. Lower panel: Temperature data are disturbed
    by normally distributed noise with a standard deviation of $\sigma
    = 0.1$~K. A higher $\epsilon$ of 0.6 is used in the inversion.}
  \label{fig:mill_inv_cut_nonoise}
\end{figure}
These considerations are only valid for the best available data. In
reality, data are noisy and boreholes are not sufficiently deep to
allow extraction of this information from the temperature data.
Figure~\ref{fig:mill_inv_cut_nonoise} compares inversion results where
only shorter logs are available and noise is added to the data. For
the case without noise, the average temperature of the last millennium
can be resolved for logs that run deeper than 300 to 500 m. For
shallower logs a strong decrease in amplitude of the GST
reconstruction is apparent.  If noise is present in the data even the
1000~m temperature is not able to fully recover the mean temperature
of the last 1000 years.  These observations are different from the
results for glacial/post-glacial reconstructions.  The reason is that
the amplitude of the transient temperature signal due to temperature
changes in the last few thousand years is one order of magnitude less
than that observed for the post-ice age temperature rise (see
figure~\ref{fig:timescales}). The input signal used here is a global
SAT reconstruction. At a specific site the amplitude of the transient
signal can be higher and thus improve the signal to noise ratio that
we assume in our example. Thus, in practice the expected signal should
be compared to the noise present in the data to assess if the desired
signal can be extracted from the data.

\begin{figure}
  \centering
  \includegraphics[width=.7\textwidth]{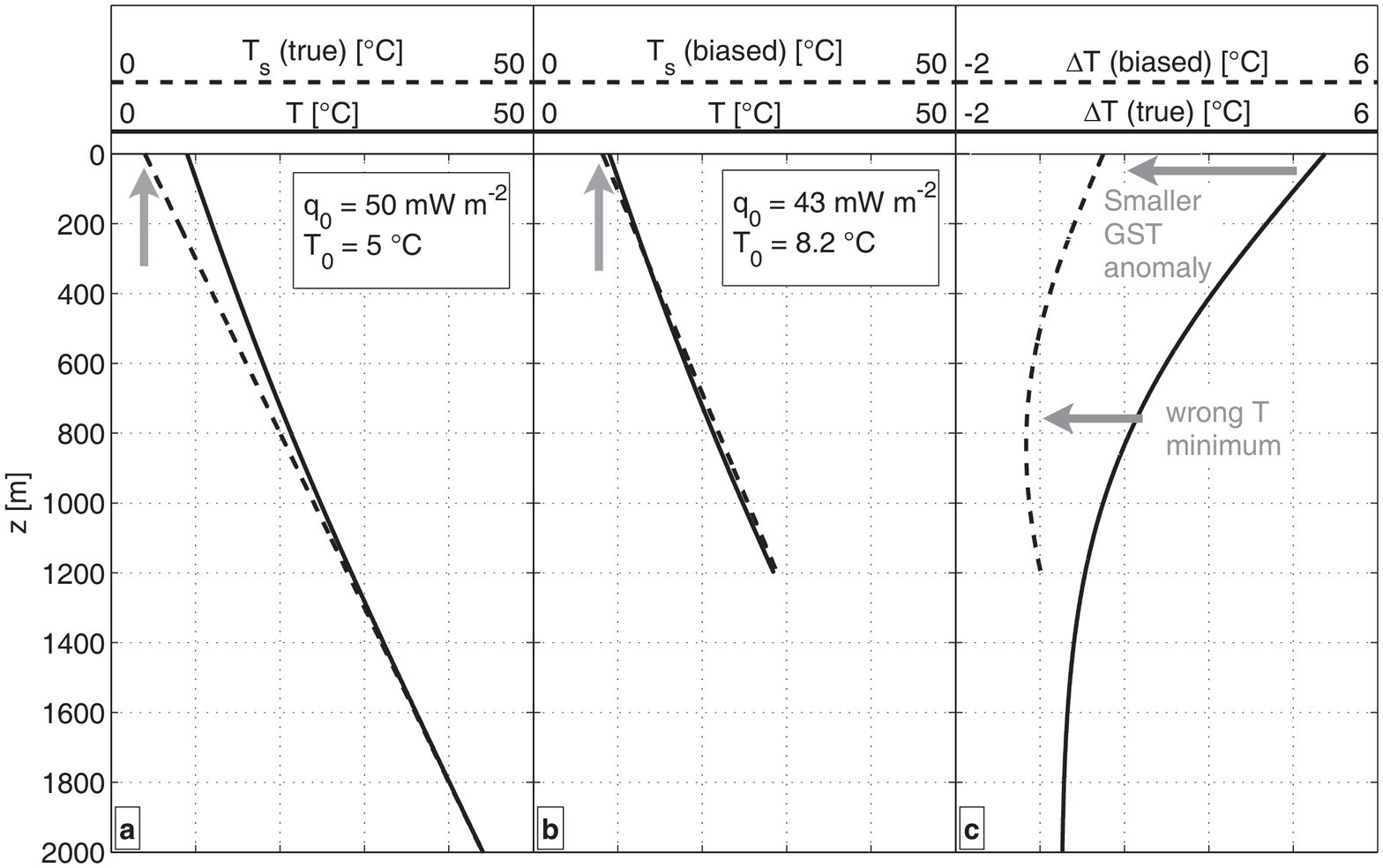}
  \caption{Estimating the transient temperature anomaly and
    undisturbed gradient results in a biased estimate for the gradient
    and a smaller temperature anomaly.}
  \label{fig:baselineproblem}
\end{figure}
Both examples show a strong dependence of the inverted GST history on
data quality and both fail to reproduce the input forcing under
certain circumstances. The ill-posed nature of the problem together
with the regularization using the damping parameter $\epsilon$ cause
this behaviour. Besides preventing numerical instability in the
inversion the procedure also minimizes the norm of the parameter
vector. Usually this effect is desired to avoid unnecessary
complicated models \citep{Constable1987}.  However, in our case a
minimal solution in this sense will have smaller GST variations
compared to the GST variations that created the transient temperature
anomaly. This is especially true since the mean GST $T_0$ and the
surface heat flow $q_0$ are also solved for.
Figure~\ref{fig:baselineproblem} illustrates this problem. The black
line denotes the true temperature perturbation due to a step increase
in ground surface temperature. If one inverts simultaneously for GST
history, $q_0$, and $T_0$ using a minimum norm model in the presence
of noise, a steady state temperature profile (grey line) tilted toward
the temperature anomaly will be obtained instead the true one (dashed
black line). In a case with a step increase in GST, the estimate for
$q_0$ will be too low and that for $T_0$ will be too high. An example
of this trend will be given in the case example. It is also notable
that the temperature perturbation crosses the steady-state profile at
two points (arrows). Thus, a false GST history is created with a
decrease in ground surface temperature and a subsequent increase
although the input model uses only a step increase. Even with an
$\epsilon$ chosen optimal for the inversion, this problem will persist
as it is imminent in the regularization.

\subsection{Thermal parameters}
\label{sec:SensThermalParmaeters}

\begin{figure}
  \begin{center}
    \includegraphics[width=\fsiz,clip,trim= 0 20 0 0]
    {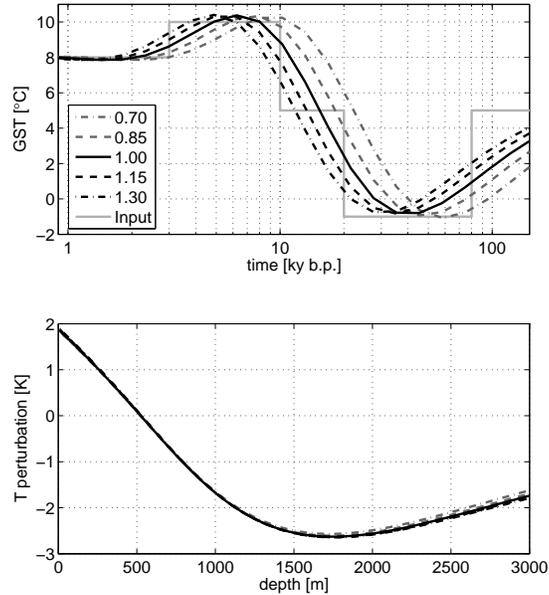}
    \caption[Inversion results for systematic changes in thermal
    diffusivity]
    {Systematic changes in thermal diffusivity cause shifts in the
      timing of the events. Thermal conductivity is varied from 70
      \%{} to 130 \%{} of the original value.  For better visibility,
      inverted curves are stacked and plotted as smooth lines rather
      than a series of step functions.}
    \label{fig:SensThermalDiff}
  \end{center}
\end{figure}
\begin{figure}
  \begin{center}
    \includegraphics[width=\fsiz{},clip,trim= 0 20 0 0]
    {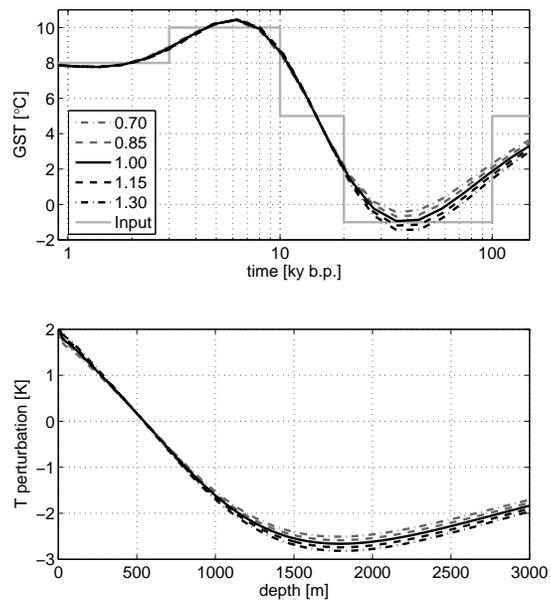}
    \caption[Inversion results for systematic changes in heat
    production]{Systematic changes in heat production rate cause
      variations in the amplitude of the events. Heat production rate
      is varied from 70 \%{} to 130 \%{} of the original value. The
      inverted curves are plotted as smooth lines rather than a series
      of step functions for better visibility of the effect.}
    \label{fig:SensHeatProd}
  \end{center}
\end{figure}

The inversion procedure requires that thermal conductivity, thermal
diffusivity, and heat production rate are exactly known. In a real
situation errors for this parameters can be minimized but cannot be
avoided altogether. For instance errors might be introduced by
preferential sampling of units that are not representative of the
whole sequence. Also, missing or false corrections for in-situ
conditions of temperature and pressure can cause systematic deviations
from the correct thermal parameters. In the following the
uncertainties in the inversion resulting from these errors are
considered.

In general, the inversion procedure interprets curvature in the
temperature log as a transient signal. Thus, only the terms
introducing curvature in equation~\ref{eq:homogenousforwardmodel}
will have an influence on the inversion result. The transient term
of the model has thermal diffusivity as its only relevant parameter.
We can expect that this property will have the largest influence on
the solution. Within the steady state terms, only the heat
production term introduces curvature. As heat production rate is
usually small, the influence of this term is expected to be less
important for boreholes of moderate depth
(see~\ref{sec:SensThermalParmaeters}). Heat production and thermal
conductivity are related inverse proportional in this term. Thus, to
analyze the errors introduced by this term, only on of the
parameters needs to be studied. We chose the heat production rate.
Thermal conductivity has its main influence in the steady-state
terms for heat conduction. A false assumption will therefore lead
directly to a proportional error in the heat flow density but does
not affect the transient part of the temperature log. In practice,
of course, thermal diffusivity is indirectly determined as the ratio
of thermal conductivity and volumetric heat capacity and will have
an indirect effect on the solution.

In order to support this qualitative discussion with quantitative
results a synthetic temperature log was inverted with false
assumptions of thermal diffusivity and heat production rate in two
separate inversion runs. The parameters were varied by $\pm$30\%{} of
the original value. A simplified GST history was used in the forward
model to compute the transient temperature anomaly. The GST history
used is comprised of several step changes that aim to reproduce the
general magnitude and timing of the expected effects and do not
represent an actual GST history. The results of these experiments are
shown in figures~\ref{fig:SensThermalDiff} and \ref{fig:SensHeatProd}.
Thermal diffusivity influences the timing of events in the
reconstructed time series. A phase shift of the entire reconstruction
occurs when a false estimate is used in the inversion.  The amplitude
of computed past temperature remains unchanged.  Heat production
influences the magnitude of the paleoclimatic events but retains
timing of events.  The effect of the heat production increases with
time, consistent with the fact that the heat production term increases
quadratically with depth. It is apparent, that the heat production
will only be significant for very large deviations from the correct
value and deep boreholes. For instance, it should be irrelevant for
millennial scale studies. Thermal diffusivity can change the timing of
events on all timescales and it is also important for inversions of
much shorter timescales than considered here.

\subsection{Heterogeneity}
\label{sec:SensHetero}

The previous discussion considered a homogeneous subsurface only.  The
question is of course whether the conclusions drawn a valid for a
heterogeneous medium. To address this question we will assess the
impact of a heterogeneous 1D layered earth model on the results of the
inversion. In the following the transient profiles computed by a FD
algorithm \citep{RathSubmitted2005} will be considered as ground
truth. Random variations of the thermal properties on distinct length
scales will be introduced and the inversion results studied.  We will
discuss the effect of systematic variations on a synthetic example of
a sedimentary rock column. We will study the steady state $T$-$z$
profile that results from compaction and elevated temperatures. 

For the forward computation a simplified GST history of the last
100\,000 years is used, with a pre-observational mean of 5°C. The
profile is computed down to 10\,000 m with a vertical resolution of
5m. For the analysis only the top 3000 m are used, the deeper parts
are only necessary to avoid edge effects in the solution. Thermal
parameters are those of table~\ref{tab:thermal_params} if not
explicitly specified.

Obviously, the largest influence on the temperature profile will come
from variations in thermal conductivity. Deviations from the straight
line that is assumed for the stationary temperature will be falsely
interpreted as noise or paleoclimate. In the homogeneous subsurface
framework it is impossible to study variations of $\lambda$ and
$\kappa$ separately. One way to overcome this problem is to cast the
steady state part of the equation in terms of the Bullard depth $z_b$,
for the homogeneous case defined by $z_b = z/\lambda$. Consider the
differential equation for 1D heat transport in a homogeneous medium:
\begin{equation}
  \label{eq:1d_trans_real_depth}
    \frac{\partial^2 T}{\partial z^2} =
    \frac{1}{\kappa}\frac{\partial T}{\partial t} + \frac{A}{\lambda}.
\end{equation}
Substituting $\partial z=\lambda\partial z_b$ yields
\begin{eqnarray}
  \label{eq:1d_trans_bullard_depth}
  \frac{\partial^2 T}{\lambda^2 \partial z_b^2}
  &=&  \frac{1}{\kappa} \frac{\partial T}{\partial t} + \frac{A}{\lambda} \\
  \Rightarrow
  \frac{\partial^2 T} {\partial  z_b^2}
  &=&  \frac{\lambda^2}{\kappa} \frac{\partial T}{\partial t} + \lambda A
\end{eqnarray}
Thus if $\kappa'= \kappa/\lambda^2$ and $A'=A/\lambda^2$,
equation~\ref{eq:homogenousforwardmodel} can be written as:
\begin{eqnarray}
  \label{eq:heterogeneous_forward_bullard}
  T(z_b) &=& T_0 + q_0 z_b + \frac{A'z_b^2}{2}  \nonumber\\
  &+& \sum\limits_{j = 1}^N T_n^G \left(
    \mbox{erfc} \left(
      \frac{z}{2\sqrt \kappa' t_j}
    \right)
    - \mbox{erfc}\left(
      \frac{z}{2\sqrt \kappa' t_{j - 1} }
    \right)
  \right)
\end{eqnarray}
Using the transformed variables and taking the heat production term
to the left hand side, the steady state part of the profile can be
reduced to a homogeneous problem. With this construct it is possible
to use a layered model for the thermal conductivity and a
homogeneous one for the thermal diffusivity. This allows us to study
the influence of thermal diffusivity alone in the algorithm. With
the general framework established, we will now turn to the different
aspects of heterogeneity.

\begin{figure}[htbp]
  \centering
  \includegraphics[width=.6\textwidth]{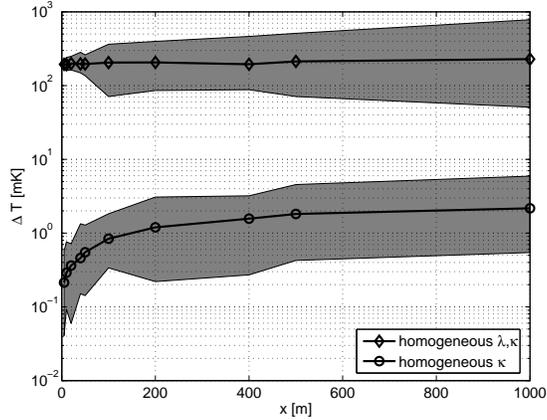}
  \caption{Misfit between temperature profiles computed from the
    FD-model and homogeneous models for varying length scales of
    variability. Mean RMS-values are given for a homogeneous
    subsurface ($\diamond$) and for homogeneous $\kappa$ only
    ($\circ$). Grey areas denote min/max range.}
  \label{fig:het_multi_len1}
\end{figure}
The noise on varying length scales has been studied by modifying the
FD model to compute the temperature profiles. The layers of the model
were assigned random values of thermal conductivity with a mean of 2.5
\uwlf{} and a standard deviation of 0.5 \uwlf{}. Volumetric heat
capacity was held constant throughout the experiment, resulting in a
simultaneous variation of thermal conductivity and diffusivity.  Ten
different configurations were considered with layer widths $x$
increasing from 5 m to 1000 m. For each configuration 50 realizations
of the thermal conductivity-depth profile were generated. These were
then used in the forward model to compute 50 temperature depth
profiles.  Additionally, temperature profile for an equivalent
half space and for a layered representation
(equation~\ref{eq:heterogeneous_forward_bullard}) were computed. The
differences between the ``true'' temperature profile and the two
profiles with assumptions are shown in
figure~\ref{fig:het_multi_len1}. The misfit of the equivalent
homogeneous model is large because variations in the steady state
temperature profile are not reflected in the homogeneous model. This
highlights the importance of the knowledge of thermal conductivity.
On the other hand, for the model with homogeneous $\kappa$ only
deviations of the order of a few mK occur, suggesting that random
variations of thermal diffusivity are smoothed out.

\begin{figure}[htbp]
  \centering
  \includegraphics[width=\textwidth{}]{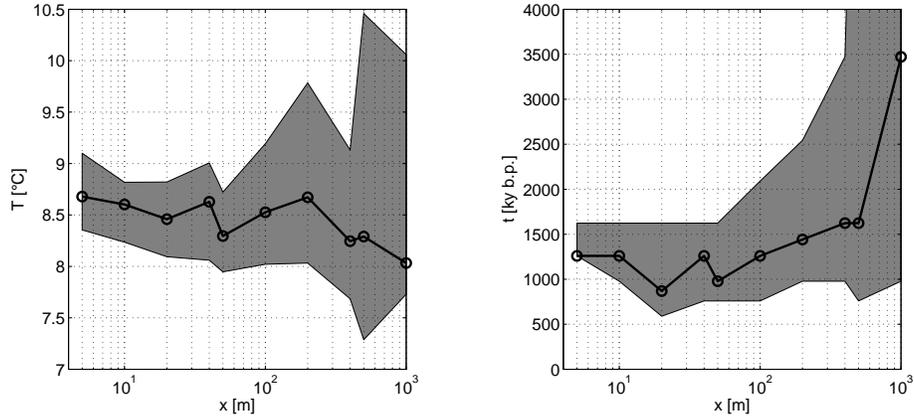}
  \caption{Variation of the timing and amplitude of the maximum
    temperature in the Holocene versus the length scale of the random
    noise as defined in the text. \emph{Left:} Value of the maximum
    temperature. \emph{Right:} Timing of the maximum value. Grey area
    denotes 75 \%{} quartiles of all realizations.}
  \label{fig:het_bmax}
\end{figure}
Next, the random data have been used as input to the inversion
procedure. For each inversion the value and timing of the maximum
temperature in the Holocene have been recorded in order to compress
the results (figure~\ref{fig:het_bmax}). To accommodate the increasing
misfit between model and data for increasing noise length scale, the
damping parameter $\epsilon$ was increased from 0.35 to 1 according to
an increase in noise length scale from 5 m to 1000 m.  This leads to a
stronger damping and is reflected in the decrease of the maximum
temperature value because models with smaller norms are preferred for
large $\epsilon$. The influence of noise on the result becomes most
pronounced for length scales larger than 100 m when the scale of the
signal is on the order of the noise scale. The timing of the event is
modified in a similar way.

\begin{figure}[htbp]
  \centering
  \includegraphics[width=\textwidth]{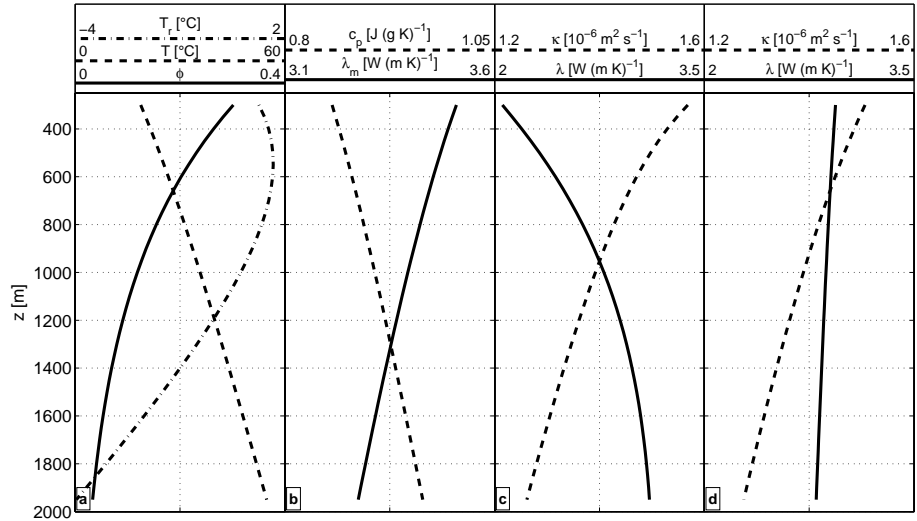}
  \caption{Systematic variation of thermal properties with depth. a)
    Porosity, temperature, and reduced temperature; b) thermal
    conductivity and heat capacity of the rock matrix, depending on
    temperature; c) effective thermal conductivity and diffusivity
    with porosities from panel a); d) Effective thermal properties for
  constant porosity of 0.1.}
  \label{fig:het_prop_compac}
\end{figure}
In crystalline rocks random variations of thermal properties might
prevail, in sedimentary rocks systematic changes are much more
important. Changing depositional environments throughout time create
diverse rock types. The influence on the inversion depends on the
particular setting. One aspect that is often neglected is the
variation of thermal properties in uniform sequences of rock due to
the temperature dependence of thermal properties and the compaction
with increased overburden. In figure~\ref{fig:het_prop_compac} a
geologic profile of a single rock type is considered. The thermal
conductivity of the matrix is taken as 3.5 \uwlf{}. An exponential
decrease of porosity due to compaction is assumed (Athy's law) with a
surface porosity of 0.45 and a characteristic depth of 750 m
\citep{Allen1990}. Thermal conductivity of the matrix is taken
inversely proportional to temperature \citep{Haenel1988} and heat
capacity of the matrix is approximately proportional to temperature in
the range considered. For simplicity the pore fluid is assumed to be
fresh water, thermal properties are taken from reference tables
\citep{Wagner2002,Kaye1968}.  Temperature and heat flow at the surface
are 10 °C and 0.06 \uhf{}, respectively. These assumptions lead to a
curvature in the thermal properties (figure~\ref{fig:het_prop_compac}
c). The curvature is reflected in the reduced temperature $T_r$
(figure~\ref{fig:het_prop_compac} a).  This effect is similar in
magnitude to the influence of paleoclimate. It is apparent that these
effects need to be taken into account for any paleoclimate inversion.

For comparison, effective thermal rock properties are also given for a
constant porosity of 0.1 to consider only the temperature/pressure
effects (figure~\ref{fig:het_prop_compac} d). The shape of the
$\lambda-z$ profile changes considerably displaying the prominent
influence of the porosity decrease. The $\kappa-z$ profiles show that
the porosity effect is compensating the temperature effect.
Diffusivity varies by about 20 \%{} in this example. When comparing
these results to the inversions of the previous sections it is clear
that thermal conductivity needs to be known accurately to achieve
reliable results. Diffusivity is again somewhat less important.

\section{Case study: KTB}

\begin{figure}
  \centering
  \includegraphics[width=\textwidth{}]{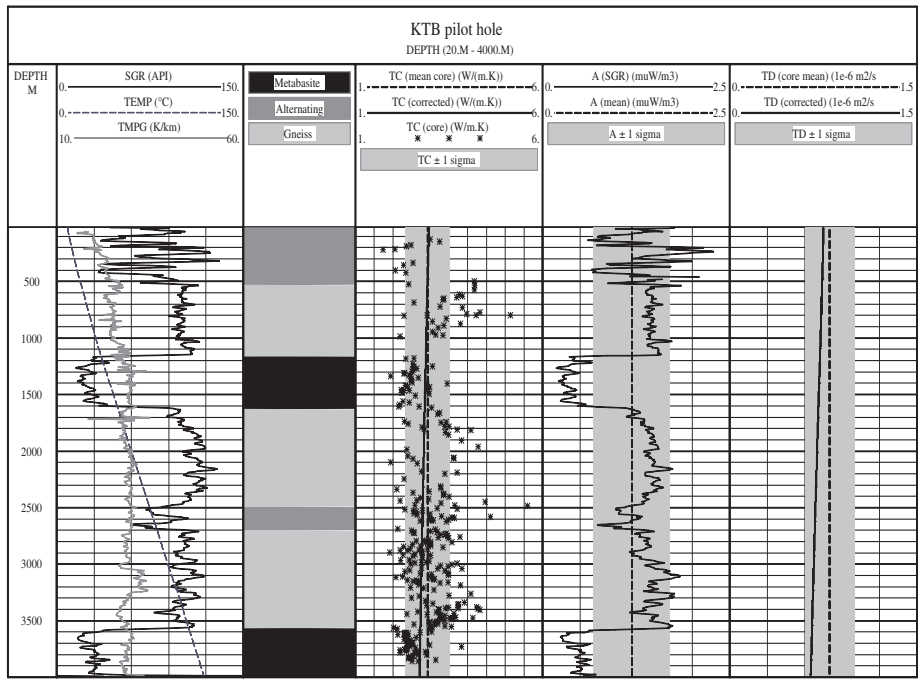}
  \caption{Composite log, KTB pilot hole. Panels from left to right: 1)
    Spectral Gamma Ray (SGR), temperature (TEMP), temperature gradient
    (TMPG); 2) Generalized lithology; 3) Thermal conductivity: Core
    measurements (TC), mean value (mean TC), $\pm 1\sigma$ of the mean
    value (shaded area), corrected for temperature and pressure (corr
    TC); 4) Heat production $A$ from spectral gamma ray (A), mean
    value of $A$ (mean A), thermal diffusivity $\kappa$ (TD).}
  \label{fig:ktb_composite}
\end{figure}

A temperature log assumed to be in equilibrium was recorded in the KTB
pilot hole down to 3990~m on September 17, 1997 and analyzed in
\citet{Clauser1999}. An earlier recording, dating from February 1996,
was analyzed in detail for thermal processes, including transient
effects of paleoclimate \citep{Clauser1997}. Shallow temperature logs
in the vicinity of the KTB site have also been interpreted in terms of
paleoclimatic influence \citep{Clauser1995a,Clauser1999}.  In the
light of the previous sections, we will revisit the analysis of the
KTB borehole and discuss the uncertainties connected with the
interpretation. In that study the temperature data was inverted for
paleoclimate in the period from 10\up{5} to 10\up{2} years before
present, using 100 time steps, a thermal conductivity of 2.92 W (m
K)\up{-1}, a thermal diffusivity of 10\up{-6} m\up{2} s\up{-1}, and a
heat production rate of 1.1 $\mu$W m\up{-3}. Their analysis yielded a
peak to peak amplitude of 10 K for the temperature increase from the
latest glacial stage to the Holocene.

The geological profile in the KTB pilot hole is primarily composed
from two lithologies: Metabasites and Gneisses. These can easily be
separated using the Gamma-ray-log (SGR)
(figure~\ref{fig:ktb_composite}, left panel). Thermal conductivity
(figure~\ref{fig:ktb_composite}, right panel) does not seem to vary
systematically over the depth of the borehole, justifying the
assumption of a homogeneous half-space. The mean thermal
conductivity is $(2.93\pm0.60)$\uwlf{}.  This value reduces to
2.82\uwlf{} if a temperature-pressure correction is applied to the
data \citep{Buntebarth1991}.  Knowledge of the thermal diffusivity
is much more uncertain. Literature cites ranges for Gneisses and
Amphibolites as 0.5 to 1.2$\cdot10^{-6}$\utd{} and 0.6 to
0.8$\cdot10^{-6}$\utd{}, respectively \citep{Landolt-V1a1982}.
\cite{Seipold1995} reports $(0.8\pm0.2)\cdot10^{-6}$\utd{} for
Amphibolite samples from neighbouring sites.  Again, a correction
for temperature and pressure of the value given by
\cite{Seipold1995} reduces the mean diffusivity to a value of
0.70\utd{}.


\begin{figure}
  \centering
  \includegraphics[width=.5\columnwidth]{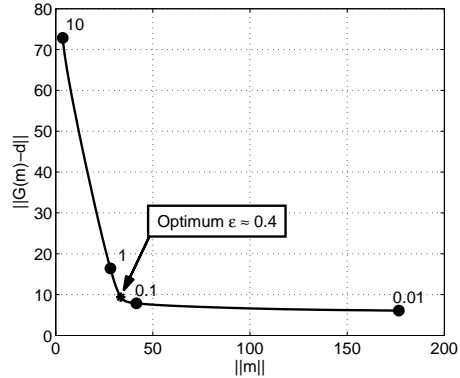}
  \caption{L-curve for the inversion of the KTB pilot hole-temperature log.}
  \label{fig:ktb_lcurve}
\end{figure}
\begin{figure}
  \centering
  \includegraphics[width=\textwidth]{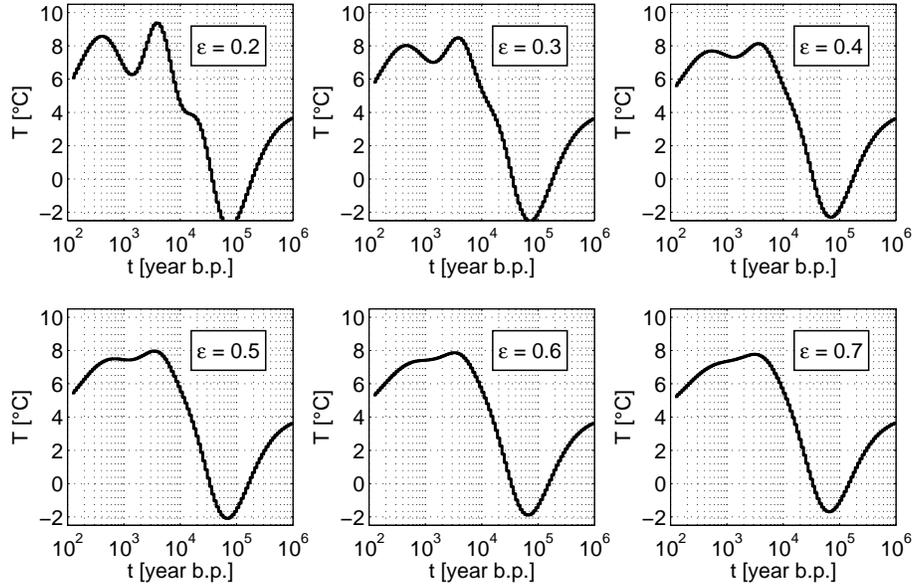}
  \caption{Inverted GST histories for varying values of the damping parameter.}
  \label{fig:ktb_gst_panel}
\end{figure}
The next step in the analysis is the choice of the optimum
regularization parameter $\epsilon$.  Figure~\ref{fig:ktb_lcurve}
shows the L-curve for varying values of $\epsilon$. The optimum
choice seems to be $\epsilon = 0.4$.  Figure~\ref{fig:ktb_gst_panel}
highlights this in more detail. The inverted time series for the
different values of $\epsilon$ are shown.  The direct comparison of
results is especially helpful to analyze which features of the GST
history are numerical oscillations and which ones represent actual
information. At a level of 0.4 oscillations that change the warming
amplitude seem to be dampened out. A more conservative interpreter
might choose an even higher value of 0.6, but in this range of
$\epsilon$-values the main features are only changed by a few tenth
of a degree Celsius.

\begin{figure}
  \begin{center}
    \includegraphics[width=\fsiz]{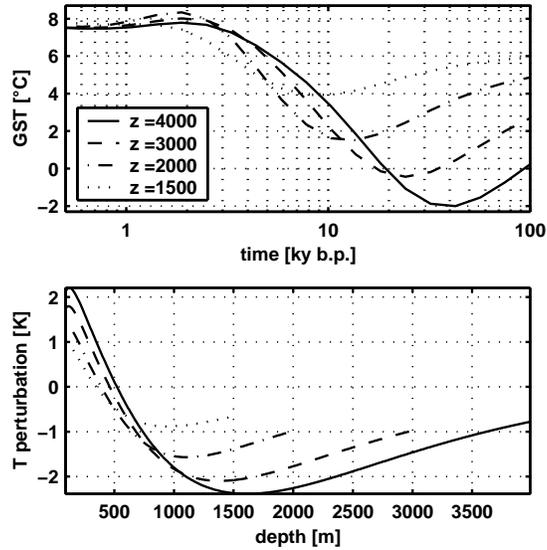}
    \caption[Inversion results of the KTB log for different
    depths]{Inversion results for the KTB log for different depths of
      the log. Top: Inverted GST history for different log length. For
      clarity, the step amplitudes of the GST histories are shown as
      continuous lines. Bottom: Transient temperature signal in the
      subsurface.}
    \label{fig:KTBCutOff}
  \end{center}
\end{figure}
Following the choice of the damping parameter, we shortened the log to
illustrate the detrimental effect of the log length.  The maximum
depth of originally 3990~m was successively cut to lengths of 3000,
2000, and 1500~m.  Figure~\ref{fig:KTBCutOff} shows the reconstructed
GSTH and transient temperature perturbation in the borehole for
various cut off depths. Deviations from the expected results become
significant for depths less than 2500~m. It is noteworthy that not
only the temperature drop of the last glacial stage decreases but also
the timing of the optimum temperature in the Holocene changes. An
interpreter of paleoclimatic inversions must be aware that the post
glacial temperature increase will influence the shape and timing of
later events as well.

\begin{table}
  \caption[Inversion results for the KTB log, cut off at various
  depths]{Inversion results for the KTB log, cut off at various
    depths. The inverted parameters vary systematically with the
    depth of the log.}
  \label{tab:KTBCutOff}
  \begin{center}
    \begin{tabular}[t]{@{}llll} \br
      Depth (m) & $T_0$  (°C)& $q$ (mW m\up{-2})&
      $dT/dz$ (mK m\up{-1}) \\\mr
      4000 & 4.14 & 86.3 & 29.6 \\
      3000 & 4.70 & 84.8 & 29.6 \\
      2000 & 5.51 & 81.8 & 28.0 \\
      1500 & 6.15 & 78.1 & 26.7 \\ \br
    \end{tabular}
  \end{center}
\end{table}
Table~\ref{tab:KTBCutOff} shows the inversion results for the
undisturbed GST (i.~e.\ the pre-observational mean), the heat flow
into the model and the undisturbed gradient. This highlights that the
systematic variation of the GST history is accompanied by an according
change in the steady-state parameters. As discussed in the section on
length of temperature logs, the heat flow value decreases whereas the
GST increases. This is consistent with inverting a step increase in
ground surface temperature using a log that is too short too fully
resolve the GST history.  It can be expected from this data that
crustal heat flow estimates may be too low by about 8 mW m\up{-2}, if
a transient signal of the last ice age has not been properly removed.

\section{Conclusions}
\label{sec:conclusion}

The sensitivity analyzes in the previous sections have shown that the
inversion results are strongly dependent on data quality. The
parameters most important depend on the time scale of inversion. For
inferences about the post-glacial temperature rise the maximum depth
of the temperature log is the most challenging requirement. If it is
met, the inversion is robust against noise in the temperature data. A
depth of more than 3000~m seems to be optimal to resolve the
temperature rise adequately. Unfortunately, undisturbed temperature
logs running that deep are rare.  Reconstructions of the last
millennium require only modestly deep logs. A depth of about 300~m to
500~m is suitable to derive a mean temperature for the last 1000
years. The exact values will depend on the magnitude of the noise and
the transient signal at a specific site. For instance the IHFC
database for borehole temperatures and climate reconstruction lists
473 out of 754 temperature logs with a maximum depth greater than
300~m (\verb#http://www.geo.lsa.umich.edu/~climate#). However, for
this type of inversion the demand on data fidelity is very high. Even
a noise level of 0.1~K seriously degrades the inversion results. In
practice the noise level a

Errors in the thermal properties, namely thermal diffusivity and heat
production, seem not to be the main source of error. The diffusive
nature of the signal and the spreading over a wide depth range greatly
dampen the influence of the thermal diffusivity. The importance of
heat production increases with reconstruction length. But even for
studies of the last 100\,000 years errors introduced by false
assumptions of the heat production rate will be small compared to
other sources of error.

The studies on the regularization parameter show that a thorough noise
analysis should accompany any attempt to invert GST histories from
geothermal data. The correct choice of the regularization is crucial
and should be justified by some objective means. A non-optimal choice
of the regularization can easily lead to artefacts that are much
larger than the calculated model errors.  Especially to small an
$\epsilon$ introduces events that could be mistaken for variations of
paleo-temperatures. We used the L-curve criterion but other methods
like generalized cross validation \cite[e.~g.,\
][]{Hansen1998,Haber2000,Aster2004} or truncated iterative methods
\cite[e.~g.,\ ][]{Hansen1998,Hanke1995,Berglund2002}could also be
utilized.

Finally, although the least-squares fit with a Tikhonov
regularization minimizing the model norm is a generally applied
method in solving ill-posed problems, it might not necessarily be
the best choice, as pointed out by \citet{Hansen1998}. Other
regularization methods might better preserve the information
contained in borehole temperatures. These include Bayesian-type
smoothing approaches \citep{Serban2001} as well as nonlinear minimum
support regularizers
\citep{Portniaguine1999,Zhdanov2002,Zhdanov2004}. Research in the
development of novel approaches to regularization and the
incorporation of prior knowledge are an urgent task for the future.

\section*{Acknowledgments}
 Our research was funded by the German Federal Environmental
Ministry as part of the activities of its AkEnd Working Group
(http://www.akend.de) through Bundesamt f\"{u}r Strahlenschutz (BFS,
Federal Agency for Radiation Protection), contract no.~9X0009-8390-0
to RWTH Aachen and contract no.~WS~0009-8497-2 to Geophysica
Beratungsgesellschaft mbH. Two anonymous reviewers provided us with
helpful suggestions.

\bibliography{pclimate}
\end{document}